%% file: main.tex
\documentclass[conference]{IEEEtran}

\usepackage[utf8]{inputenc}
\usepackage[T1]{fontenc}
\usepackage[swedish,english]{babel}

\usepackage{graphicx, float, subfigure, blindtext, caption}

\usepackage{tikz,xcolor,hyperref}

\definecolor{lime}{HTML}{A6CE39}
\DeclareRobustCommand{\orcidicon}{%
	\begin{tikzpicture}
	\draw[lime, fill=lime] (0,0) 
	circle [radius=0.16] 
	node[white] {{\fontfamily{qag}\selectfont \tiny ID}};
	\draw[white, fill=white] (-0.0625,0.095) 
	circle [radius=0.007];
	\end{tikzpicture}
	\hspace{-2mm}
}

\foreach \x in {A, ..., Z}{%
	\expandafter\xdef\csname orcid\x\endcsname{\noexpand\href{https://orcid.org/\csname orcidauthor\x\endcsname}{\noexpand\orcidicon}}
}

\usepackage{hyperref}
\hypersetup{
    colorlinks=true,
    linkcolor=blue,
    filecolor=blue,      
    urlcolor=blue,
    pdfpagemode=FullScreen,
    }
\usepackage{mathtools}
\usepackage{cite}
\usepackage[nolist,nohyperlinks]{acronym}
\usepackage{cleveref}
\graphicspath{{images/}}

\usepackage{titlesec}
\titlespacing*{\section}{0pt}{*1}{*1}
\titlespacing*{\subsection}{0pt}{*1}{*1}
\renewcommand{\thesubsubsection}{\arabic{subsubsection}}
\titleformat{\subsubsection}[runin]{\itshape}{\thesubsubsection)}{1em}{}[:]
\titlespacing*{\subsubsection}{\parindent}{0pt}{*1}

\author{\IEEEauthorblockN{
Manuel Striani\orcidA{} 
}
\IEEEauthorblockA{
Computer Science Institute\\
DiSIT - Department of Sciences and Technological Innovation\\
University of Piemonte Orientale, Alessandria (Italy)\\
manuel.striani@uniupo.it
}}

\title{NRTS: A Client-Server architecture for supporting data recording, transmission and evaluation of multidisciplinary teams during the neonatal resuscitation simulation scenario}
\begin{document}
\maketitle
\input{sections/abstract}
\input{sections/keywords}

\input{sections/introduction}

\input{sections/method}
\input{sections/results}

\input{sections/download}
\input{sections/acknowledgement}
\input{sections/contacts}
\bibliographystyle{IEEEtran}
\bibliography{refs.bib}
\end{document}

%% file: sections/abstract.tex
\begin{abstract}
In this technical report, we describe \textbf{N}eonatal \textbf{R}esuscitation \textbf{T}raining \textbf{S}imulator (\textbf{NRTS}), an Android mobile app designed to support medical experts to input, transmit and record data during a High-Fidelity Simulation course for neonatal resuscitation. This mobile app allows one to automatically send all the recorded data from “Neonatal Intensive Care Unit” (NICU) of Casale Monferrato Children’s Hospital, (Italy) to a server located at the Department of Science and Technological Innovation (DiSIT), University of Piemonte Orientale (Italy). Finally, the medical instructor can view statistics on a simulation exercise that may be used during the de-briefing phase for the evaluation of multidisciplinary teams involved in the simulation scenarios.

\end{abstract}

%% file: sections/keywords.tex
\begin{center}
\end{center}

%% file: sections/introduction.tex
\section{Introduction}
\label{section:intro}

The delivery room is a complex and dynamic environment, and although most new-borns present a normal adaptation to birth, about 5-10\% of new-borns, need some form of help in the transition from intra-to extra-uterine life, 3-5\% require mask ventilation, and only less than 1/1\% require advanced resuscitation measures such as intubation, chest compressions, ventilation and medication.

Therefore, in view of the relative infrequency of new-born babies requiring extensive resuscitative interventions and the complexity and dynamism of neonatal emergencies, the delivery room can become a high-risk place for patients and very stressful for the care staff, who must be fully aware of the various steps and be trained and prepared to skilfully perform the procedures and work effectively in a team. In 2004, the Joint Commission published a sentinel event alert on neonatal outcomes indicating ineffective communication within the neonatal care team as the leading cause of death or permanent disability.

Increasing efforts to reduce medical errors and risks associated with physicians’ performance of high-risk procedures. on patients have led to the increasing use of simulation in emergency medicine training. Before 2010 the Neonatal Resuscitation

Training Programme Neonatal Resuscitation Training Programme (NRP) of the American Academy of Pediatrics (AAP) was focused on acquiring the technical knowledge and skills necessary for neonatal resuscitation.

In recent years, there has been a gradual transformation to simulation-based training aimed at acquiring and maintaining the technical skills for teamwork in the management of critical patients.

By combining teaching and risk analysis in a safe environment for healthcare workers, medical simulation can decrease errors caused by human factors and create a more effective and safer patient treatment. It can only be a strategic and innovative tool in the context of multidisciplinary lifelong learning if there is a basis for the implementation of standardized high-quality programs of verifiable effectiveness. Simulation is an ideal educational methodology for teaching cognitive, technical and behavioural skills and hopefully improving early life care. In fact, it allows training that goes from a simple and linear learning of the resuscitation algorithm to a more complex one, such as behavioural skills and teamwork through debriefing, an essential part of the simulation during which, immediately after the scenarios, an analysis of the procedures is performed in search of potential weaknesses and strategies for the improvement, prevention and containment of errors.

The continuous training of healthcare workers becomes essential in the context of emergency and urgent care, where professionals are called upon to work in critical and highly stressful situations, and where the possession of manual dexterity in the execution of complex manoeuvres, prompt decision-making and the ability to work in a team become vital for the success of interventions.

This technical report presents a Client-Server architecture for using a mobile app for \textbf{N}eonatal \textbf{R}esuscitation \textbf{T}raining \textbf{S}imulator (\textbf{NRTS}), designed and developed for supporting the correctness of the neonatal resuscitation during the simulation scenario and the subsequent evaluation of multidisciplinary teams in the debriefing phase.

%% file: sections/method.tex
\section{Neonatal high Fidelity simulation center}

The neonatal HFS (High-Fidelity Simulation) centre at ``Santo Spirito" Children’s Hospital of Casale Monferrato (Italy) consists of a scenario room (shown in Figure \ref{workflowArchitecture}) with DR (delivery room) or neonatal intensive care bed, a director’s room and classrooms for theoretical lessons and de-briefings. The simulation room was modified specifically to have the appearance of a real DR or neonatal intensive care bed. Participants had all the necessary materials for attending to a new-born available, according to the latest American Heart Association (AHA) and Academy of Paediatrics (AAP) recommendations, including: an oxygen-air blender; a T-piece resuscitator Neo-Tee$^{\textcircled{R}}$ Infant T-Piece Resuscitator Mercury Medical, Clearwater, Florida, USA); respiratory support devices for invasive (Leoni Plus, Heinen Lowenstein, Rheinland-Pfalz, Germany); respiratory support devices for invasive (Leoni Plus, Heinen Lowenstein, Rheinland-Pfalz, Germany) and non-invasive (Instant Flow, CareFusion, Hoechberg, Germany) ventilation

During scenario performance, it was possible to see the patient's vital signs and laboratory or instrumental tests on a specific monitor. Scenarios were performed by using new-born simulators (SimNewB and Premature Anne\footnote{https://laerdal.com/us/products/simulation-training/obstetrics-pediatrics/premature-anne/}). SimNewB is a highly realistic neonatal simulator with one size and weight of a new-born baby girl delivered at term with approximately 3.5 Kg body weight. Premature Anne is a highly realistic 25-week preterm infant simulator with an approximate weight of 0.6 Kg. A recording system with three high-definition cameras and an ambient microphone located in the resuscitation warmer was used.

The HFS courses were performed over a time period of 1 month between June and July, into two separate sections including theoretical and videos lessons, TS (technical skills) exercises, scenario performances, de-briefings and psychological tests. At the beginning of each section forty-three practitioners (obstetricians, neonatologists, physicians, midwives, and paediatric nurses) were admitted to the training High Fidelity Simulation centre and grouped into multidisciplinary teams of 3-4 persons and underwent simulator suite orientation (familiarization).

\section{Client side: the NRTS mobile app}
\label{section:ClientSide}

The client-side architecture offers the \textbf{N}eonatal \textbf{R}esuscitation \textbf{T}raining \textbf{S}imulator (\textbf{NRTS}), a mobile app which enable a medical expert to input, transmit and record data during the simulation phase of neonatal resuscitation. This mobile app automatically sends all the recorded data to a server in the form of process traces (i.e., the sequences of activities executed on the premature Anne at the hospital during the simulation scenarios).

In particular, \textbf{NRTS} supporting a medical instructor during the simulation scenarios to memorize and automatically send all the process traces to a server physically located at the Department of Science and Technological Innovation (DiSIT), University of Piemonte Orientale.

Furthermore, the medical instruction is also enable to insert otyer significant patient and simulation data like body-temperature or other kinds of qualitative notes and comments on teams involved in the simulation session.

Furthermore, the medical instructor may also enter other significant patient and simulation data like body temperature or other kinds of qualitative notes and comments on teams involved in the simulation session.

The mobile app \textbf{NRTS} has been developed and designed to runs on smartphone devices with a minimum operating system of Android v.13 ``Tiramisù".

\begin{figure}
\centering
\includegraphics[scale=0.5]{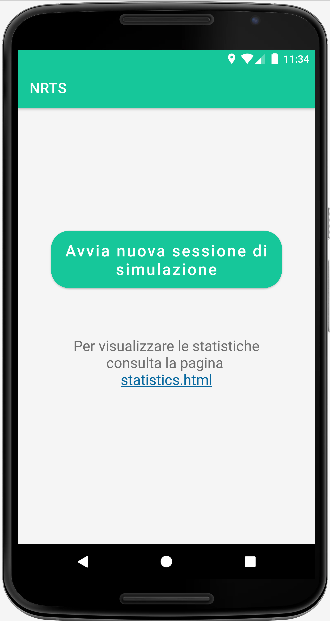}
\caption{By using this first activity, the instructor can start a new simulation session for neonatal resuscitation. In addition, the initial activity contains a link to the web page that shows the instructor, the results of the calculation of specific statistics, such as the average of the distances from the gold-standard trace (guideline) that were obtained by each group that participated in a simulation session.}
\label{figura_schermata1}
\end{figure}

The mobile application presents a first step (initial activity), shown in Figure \ref{figura_schermata1}, that offers the option of starting a new simulation session. 

\begin{figure}
\centering
\includegraphics[scale=0.5]{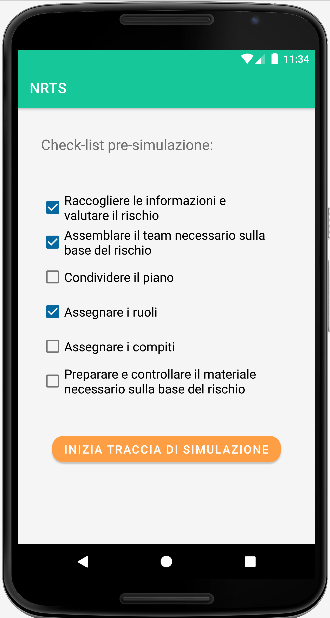}
\caption{Second activity: check-list of pre-rianimation phase}
\label{figura_schermata2}
\end{figure}

After selecting the option ``Start a new simulation session", the medical instructor, as shown in Figure \ref{figura_schermata2}, is able to visualize the checklist of pre-resuscitation actions to be performed, according to clinical guidelines, and then, it would be possible to indicate whether or not these pre-requisites have been performed before starting to record the actual simulation scenario.

It is possible to start recording the simulation session (in the form of a process trace). When the physician starts a new simulation scenario, the app executes a timer (shown in Figure \ref{figura_schermata2}) within an activity that contains a set of actions to be executed during the first minute of simulation.

The duration of each individual action in the process trace is calculated by touching the same button corresponding to the same action twice, to start and stop the timer.

\begin{figure}
\centering
\includegraphics[scale=0.5]{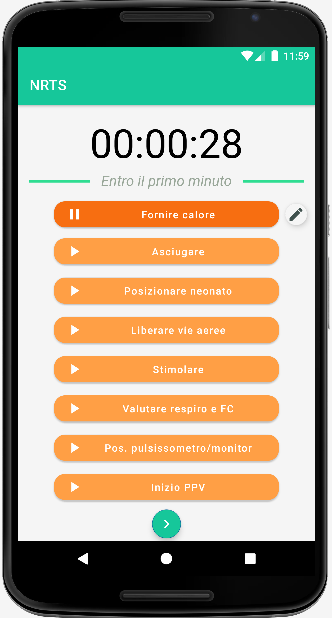}
\caption{Activity containing actions to be performed within the first minute of the simulation session}
\label{figura_schermata3}
\end{figure}

Figure \ref{figura_schermata3} shows a dialog window in which the team leader can insert the value of body temperature. The temperature ranges are decided by the domain expert, selecting temperatures from  35.5$^{\circ}$C to 39.5$^{\circ}$C and over 40$^{\circ}$C.

\begin{figure}
\centering
\includegraphics[scale=0.5]{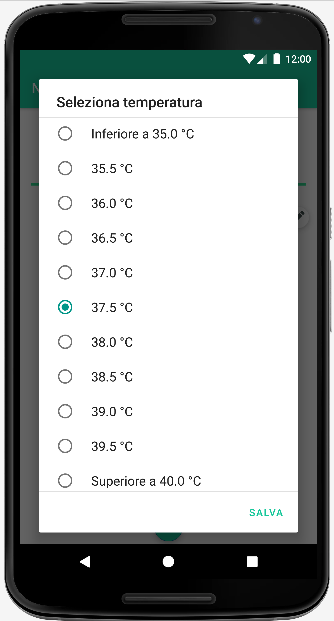}
\caption{Activity for selecting the body temperature value}
\label{figura_schermata31}
\end{figure}

\begin{figure}
     \centering
     \begin{subfigure}
         \centering
         \includegraphics[scale=0.5]{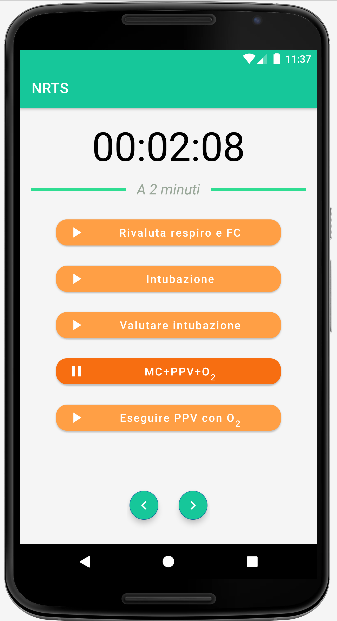}
         \caption{Activity containing actions to be performed in the the second minute of the simulation session}
         \label{figura_schermata5}
     \end{subfigure}
     \begin{subfigure}
         \centering
         \includegraphics[scale=0.5]{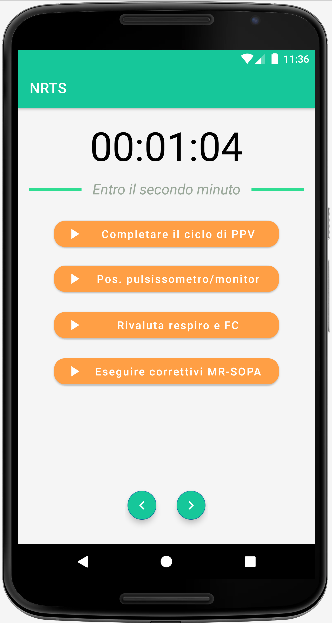}
         \caption{Activity containing actions to be performed no later than the second minute of the simulation session}
         \label{figura_schermata4}
     \end{subfigure}
\end{figure}

After performing the first minute actions, the operator switches to the screen with the actions to be undertaken within the second minute (shown in Figure \ref{figura_schermata4}). From this activity, it is possible to return to the previous activity or to proceed with the actions to be performed in the subsequent phases of the simulation scenario. Figure \ref{figura_schermata5} shows the activity with the actions to be performed within two minutes from the start of the simulation scenario.

\begin{figure}
\centering
\includegraphics[scale=0.5]{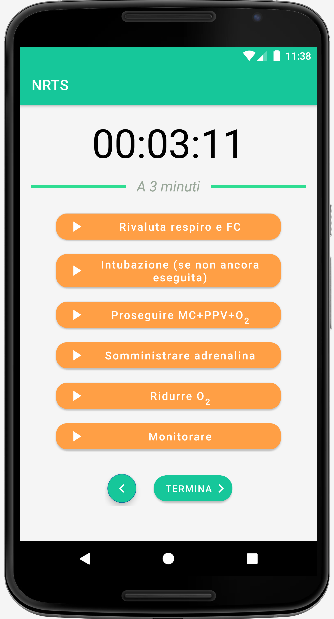}
\caption{Activity containing actions to be performed within three minutes of the simulation session.}
\label{figura_schermata6}
\end{figure}

Figure \ref{figura_schermata6} shows the activity with the actions to be performed no later than three minutes after the start of the simulation scenario. If necessary, or if the team has finished the simulation scenario, the operator may move forward to the next activity by selecting ``End". In this way, a process trace containing all information about the sequence of actions and their durations is sent to the server.

\begin{figure}
     \centering
     \begin{subfigure}
         \centering
         \includegraphics[scale=0.5]{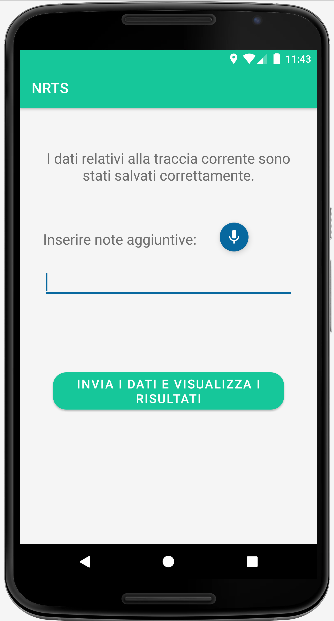}
         \caption{Via this activity, the medical instructor can send to the server additional notes or terminate the current simulation scenario}
         \label{figura_schermata7}
     \end{subfigure}
     \begin{subfigure} 
         \centering
         \includegraphics[scale=0.5]{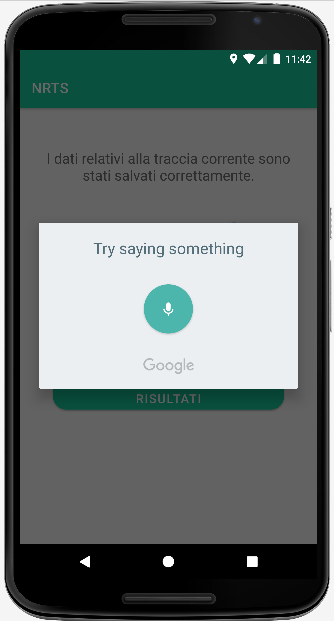}
         \caption{Activity for Speech-To-Text notes}
         \label{figura_schermata71}
     \end{subfigure}
\end{figure}

Figures \ref{figura_schermata7} and \ref{figura_schermata71} show that, before sending the process trace to the server, through the speech-to-text function, the medical instructor in charge of the simulation may enter additional notes and comments, containing information that will later be used during the de-briefing phase.

\begin{figure}[htbp]
\centering
\includegraphics[scale=0.5]{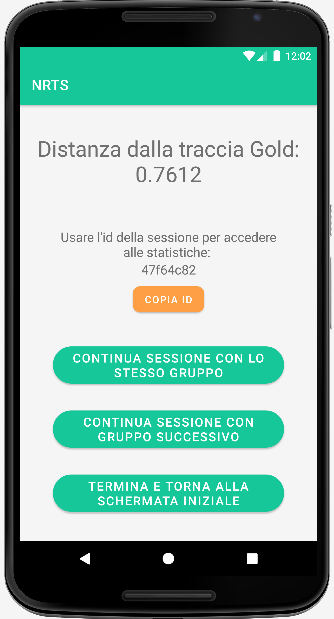}
\caption{Final activity showing the distance value between the recorded process traces during the current simulation scenario and the gold-standard trace (stored on the server). The distance of 0.7612 indicates a distance of 76\% from the guideline (gold-standard trace).}
\label{figura_schermata8}
\end{figure}

By clicking on ``Send Data", all the data collected during the simulation scenario, is saved into JSON objects and sent to the server via a HTTP-POST. After receiving the data (process trace), the server calculates the semantic distance \cite{montani-eswa15}  from the gold-standard trace (provided my medical expert) and sends the result to the mobile application. Figure \ref{figura_schermata8} shows the distance value calculated by the server, relative to the process traces recorded during the current simulation scenario compared to the gold-standard trace (guideline, based on the new-born simulation algorithm).

\begin{figure}
\centering
\includegraphics[scale=0.5]{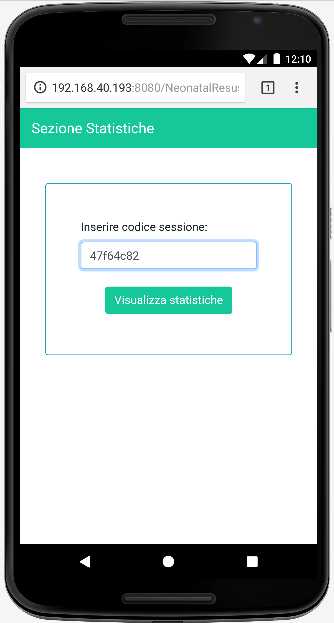}
\caption{By using the web-page, medical instructor can insert the \textsl{session ID} (provided by the \textbf{NRTS} mobile app) and visualize some statistics about different groups involved in the simulation scenario}
\label{figura_sitoWeb1}
\end{figure}

In order to view statistics for all groups that participated in the simulation session, the medical instructor is able to insert the \textsl{session ID} (provided by the \textbf{NRTS} mobile app) in the text box available on the servers-side web-page shows in Figure \ref{figura_sitoWeb1}. The activity shown in Figure \ref{figura_schermata8}, offers the possibility of automatically copy the session ID and paste it to obtain the results of all the groups that are involved in the simulation scenarios. 

\begin{figure}
\centering
\includegraphics[scale=0.5]{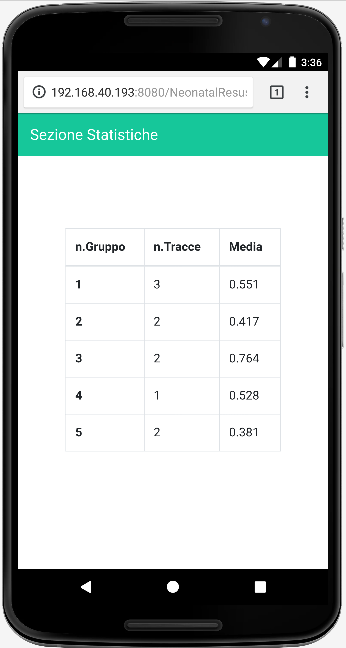}
\caption{The web page shows some statistics related to the calculation of the average distance between all recorded process traces in a simulation session, performed by the different groups.}
\label{figura_sitoWeb2}
\end{figure}

Moreover, the medical instructor is allowed to analyse groups using statistics (shown in Figure \ref{figura_sitoWeb1} and \ref{figura_sitoWeb2}) during the de-briefings phase to evaluate problems and errors that arose in the simulation session where personnel were involved (grouped into multidisciplinary teams of 3–4 persons (obstetrician, neonatologist, midwife, paediatric nurse) and underwent simulator suite orientation (familiarization).

\section{Server side}
\label{section:ServerSide}

The backend usually consists of three parts: a web server (provided by Apache HTTP or Microsoft Windows Server), a JAVA module, and a database (provided by DBMS NoSQL MongoDB\footnote{https://www.mongodb.com/}). The server component of the architecture (shown in Figure \ref{workflowArchitecture}) receives the JSON-LD data sent by the app and stores them in the NoSQL database MongoDB. In our current implementation, the server has the following properties:
\begin{itemize}
\item \textbf{Ubuntu edition}: 20.04 LTS Focal Fossa
\item \textbf{Processor}: Virtual Machine 2VCPU
\item \textbf{Installed memory (RAM)}: 4 GB
\item \textbf{Storage memory}: 40 GB
\end{itemize}

\section{The workflow architecture}
\label{workflowArchitectureSection}

\begin{figure*}[t] 
\centering
\includegraphics[width=0.9\textwidth]{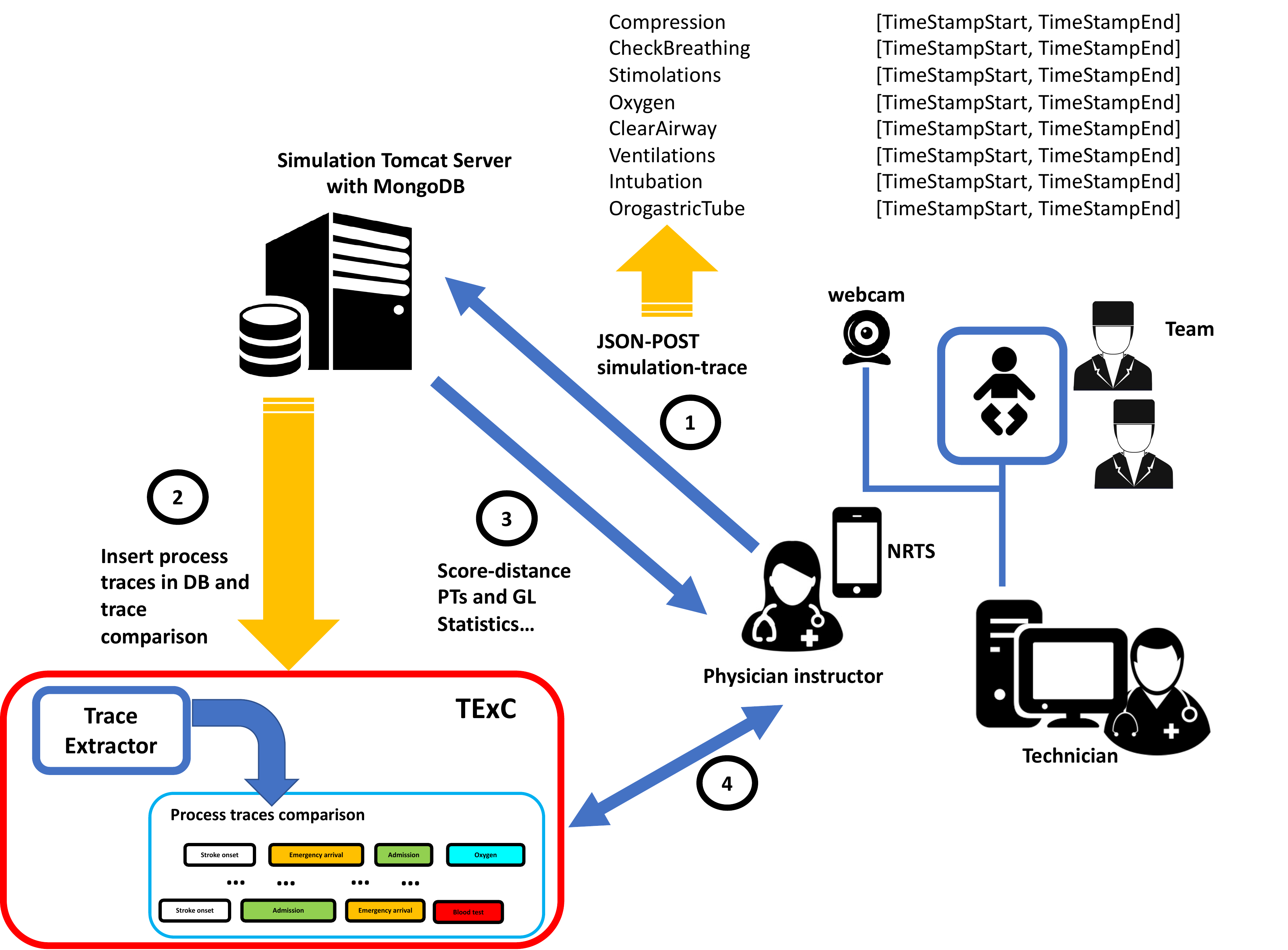}
\caption{Client-server architecture for \textbf{NRTS} Android mobile app into the HFS simulation center}
\label{workflowArchitecture}
\end{figure*}

Figure \ref{workflowArchitecture} shows the HFS simulation centre. The simulation room was composed by medical instructor and healthcare personnel grouped into multidisciplinary teams of 3–4 persons (obstetrician, neonatologist, midwife, paediatric nurse).

In the simulation room, there is a neonatal cradle with a doll representing the premature new-born infant on which a series of manoeuvres must be performed by the team. These manoeuvres are included in a pre-defined set of actions provided by the domain expert neonatologist, each of which is characterised by a start and end timestamp. In this way, each action has a specific duration. Other actions, such as fever measurement, are characterised by a range [35.5$^{\circ}$C - 39.5$^{\circ}$C] with step 0.5 provided by the medical expert. Moreover, the information on the timing and duration of each action was formalized, following the neonatal resuscitation guideline \cite{ARNOLD2011357,MADAR2021291,huang_educational_2019} describes the steps to be followed during the process of assessing and resuscitating a new-born baby.

\begin{enumerate}
\item During the simulation scenarios, the \textbf{NRTS} mobile app, through a minimalist interface, the medical instructor store process traces (sequences of actions performed by teams involved) and parameters recorded during an action (such as the body temperature value $T$ and saturation levels $SpO_2$ (shown in Figure \ref{figura_schermata31})).

\item The process trace recorded by \textbf{NRTS} mobile app, is sent to the server. The \textbf{TExC} module (shown in Figure \ref{workflowArchitecture}) calculates the distance, expressed as a value in the range [0,1] from the current process trace and gold-standard trace.

\item Finally, the server stores the process traces in a NoSQL database using the DBMS MongoDB and sends the score to the \textbf{NRTS} mobile app, which makes it available to the medical instructor, displays it in the final activity (shown in Figure \ref{figura_schermata8}).

\item The Client-Server architecture allows the storage and comparison of simulation process traces in the DBMS MongoDB, with the reference guideline for neonatal resuscitation, in order to establish the goodness and correctness of the simulation sessions with the final evaluation of the groups in the de-briefings phase.
\end{enumerate}

\section{Evaluation of multi-disciplinary teams}

The \textbf{NRTS} mobile app, makes possible to record all individual simulation sessions that are performed by the team of healthcare professionals participating in the training phase of the course. From the simulation sessions, it is possible to derive a collection of process traces, sequencing in each trace the actions performed by each individual team of participants and enriching them with temporal information (i.e. start and end timestamps).

In order to evaluate the learning outputs, it will then be possible to compare the traces generated by a team against a ``Gold" trace, which represents the correct behaviour to be adopted in the simulated situation, and thus derive an ordering of the traces generated by the various teams, with respect to their distance from the optimal trace.

The sequences of actions constituting a single trace have both temporal information attributes (i.e., start/end timestamp, and consequently duration) and atemporal attributes (such as name and measured value (body temperature $T$, oxygen $O$ or saturation $SpO_2$).

In a medical emergency domain, temporal information are very relevant and plays a central role. It may be important to penalize the fact that the same action may have different durations in different process traces. It must also be borne in mind that, for medico-legal purposes, abnormal durations relating to the same action and delays must be justified. In order to obtain an accurate calculation of the distance between two process traces, it is important to take into account all types of information that can be retrieved from the process traces themselves, therefore the simulation scenarios were evaluated both through a metric of comparison between traces, as well as with regard to times, i.e. durations were calculated as the average of the $k$-traces over the n groups that participated. The comparison of the traces was performed using the TExC (\textbf{T}race \textbf{E}xtractor and \textbf{C}omparison) module, which already existing as a Java library developed at the University of Piemonte Orientale - DiSIT, Department of Science and Technological Innovation - Computer Science institute. To calculate the distance between two process traces, we used the \textsl{Semantic Trace Edit Distance}\cite{DBLP:journals/is/MontaniL14}\cite{DBLP:journals/eswa/MontaniLSQC17}.

%% file: sections/results.tex

%% file: sections/download.tex
\section{How to download and install NRTS}

Downloading apps from the official Google Play Store is a pretty simple process. Errors like ``Not available for your device" or ``Incompatible with your device" can interrupt it. Well, these messages mean you can’t install the apps, but that doesn't mean you can't use them on your device.

Finding and downloading APK versions of those specific apps can help you install the applications manually. The most secure and exact way to download Android and install apps from Google Play, is using device with version $\geq$ Android 13 called ``Tiramisù", developed by Google, released for the public on August, 2022. \textbf{NRTS} mobile app is available on Google Play. 


For visualize statistics, our server is reachable at this IP address


\begin{center}
\url{https://tinyurl.com/NRTStatistics}{\color{blue}}
\end{center}


%% file: sections/acknowledgement.tex
\section*{Acknowledgment}
The authors would like to thank Mariachiara Martina Strozzi - medical doctor (MD) and director of the HFS center at ``Santo Spirito" Children's Hospital of Casale Monferrato (Italy) and Roberta De Benedictis, master student of Computer Science at Computer Science Institute of DiSIT - University of Piemonte Orientale (Italy), for their help and support during the process of design, developing and testing of the \textbf{NRTS} mobile architecture.
\ \\

This research is original and has a financial support of the University
of Piemonte Orientale (DiSIT-UPO).

%% file: sections/contacts.tex
\section*{Contacts}

Should you have any questions about \textbf{N}eonatal \textbf{R}esuscitation \textbf{T}raining \textbf{S}imulator (\textbf{NRTS}), do not hesitate to contact us.
For general questions you can contact dott. Manuel Striani (manuel.striani@uniupo.it), Prof. Stefania Montani (stefania.montani@uniupo.it) or Prof. Massimo Canonico (massimo.canonico@uniupo.it).